\documentclass[showpacs,aps,prl,twocolumn,superscriptaddress]{revtex4-1}
\usepackage{graphicx} 
\usepackage{dcolumn}
\usepackage{bm}
\usepackage{amssymb,amsmath}
\usepackage{epstopdf}
\usepackage{color}
\usepackage{mathrsfs}
\def\etal{$\it{et~al.}$}

\begin{document}
\title{Landau level spectroscopy of massive Dirac fermions in single-crystalline ZrTe$_5$ thin flakes}

\author{Y. Jiang}
\affiliation{School of Physics, Georgia Institute of Technology, Atlanta, Georgia 30332}
\author{Z. L. Dun}
\affiliation{Department of Physics and Astronomy, University of Tennessee, Knoxville, Tennessee 37996}
\author{H. D. Zhou}
\affiliation{Department of Physics and Astronomy, University of Tennessee, Knoxville, Tennessee 37996}
\author{Z. Lu}
\affiliation{National High Magnetic Field Laboratory, Tallahassee, Florida 32310}
\affiliation{Department of Physics, Florida State University, Tallahassee, Florida 32306}
\author{K.-W. Chen}
\affiliation{National High Magnetic Field Laboratory, Tallahassee, Florida 32310}
\affiliation{Department of Physics, Florida State University, Tallahassee, Florida 32306}
\author{S. Moon}
\affiliation{National High Magnetic Field Laboratory, Tallahassee, Florida 32310}
\affiliation{Department of Physics, Florida State University, Tallahassee, Florida 32306}
\author{T. Besara}
\affiliation{National High Magnetic Field Laboratory, Tallahassee, Florida 32310}
\author{T. M. Siegrist}
\affiliation{National High Magnetic Field Laboratory, Tallahassee, Florida 32310}
\affiliation{Department of Chemical and Biomedical Engineering, Florida A\&M University-Florida State University (FAMU-FSU)\\ College of Engineering, Florida State University, Tallahassee, Florida 32310}
\author{R. E. Baumbach}
\affiliation{National High Magnetic Field Laboratory, Tallahassee, Florida 32310}
\author{D. Smirnov}
\affiliation{National High Magnetic Field Laboratory, Tallahassee, Florida 32310}
\author{Z. Jiang}
\email{zhigang.jiang@physics.gatech.edu}
\affiliation{School of Physics, Georgia Institute of Technology, Atlanta, Georgia 30332}
\date{\today}

\begin{abstract}
We report infrared magneto-spectroscopy studies on thin crystals of an emerging Dirac material ZrTe$_5$ near the intrinsic limit. The observed structure of the Landau level transitions and zero-field infrared absorption indicate a two-dimensional Dirac-like electronic structure, similar to that in graphene but with a small relativistic mass corresponding to a 9.4 meV energy gap.  Measurements with circularly polarized light reveal a significant electron-hole asymmetry, which leads to splitting of the Landau level transitions at high magnetic fields. Our model, based on the Bernevig-Hughes-Zhang effective Hamiltonian, quantitatively explains all observed transitions, determining the values of the Fermi velocity, Dirac mass (or gap), electron-hole asymmetry, and electron and hole $g$-factors.
\end{abstract}


\maketitle
Zirconium pentatelluride (ZrTe$_5$) has long been recognized as a layered thermoelectric material \cite{thermo}. It has attracted substantial interest lately in the wave of Dirac and topological material exploration \cite{Vafek}, due to the theoretical prediction of a large-gap quantum spin Hall insulator phase in its monolayer form \cite{Thoery1_FZ}. Theory also predicts that the electronic structure of bulk ZrTe$_5$ resides near the phase boundary between weak and strong topological insulators (TIs) \cite{Thoery1_FZ,NewTheory_Zhou}, providing an ideal platform for studying topological phase transitions. Surface-sensitive spectroscopy techniques such as angle-resolved photoemission spectroscopy (ARPES) and scanning tunneling spectroscopy have recently been used to probe the surface and bulk states of ZrTe$_5$ \cite{Arpes0_GDG,ARST_SHP,STM_XQK,Arpes1_MG,Arpes4_AC,Arpes5_YLC,Arpes2_XJZ}. Intriguingly, results from different groups lead to conflicting interpretations ranging from strong/weak TI \cite{ARST_SHP,STM_XQK,Arpes1_MG,Arpes4_AC,Arpes2_XJZ} to Dirac semimetal \cite{Arpes0_GDG,Arpes5_YLC}.

On the other hand, infrared (IR) spectroscopy is a bulk-sensitive technique. Recent IR reflectance studies have suggested that bulk ZrTe$_5$ is a three-dimensional (3D) massless Dirac semimetal \cite{Refl1_NLW,Refl_NLW,ReflTP_XFX}. However, the accuracy of the transition energies extracted from the reflectance measurements may be questioned since a true Kramers-Kronig transformation cannot be implemented within the limited spectral range of magneto-IR reflectance measurements \cite{Refl_NLW,ReflTP_XFX}. Therefore, magneto-IR transmission measurements are needed to quantitatively describe the exact topological nature of ZrTe$_5$.

In this Letter, we present the IR transmission magneto-spectroscopy study of mechanically exfoliated ZrTe$_5$ thin crystals near the intrinsic limit. Because of the low carrier density, we are able to observe a series of interband Landau level (LL) transitions that exhibit the characteristic dispersion of two-dimensional (2D) massive Dirac fermions\textemdash a signature of the 2D Dirac semimetal electronic structure. We employ high-field magneto-spectroscopy with circularly polarized IR light to resolve a four-fold splitting of low-lying LL transitions, which is attributed to the combined effect of finite mass, large \textit{g}-factor, and electron-hole asymmetry.

\begin{figure}[t]
\includegraphics[width=8.5cm]{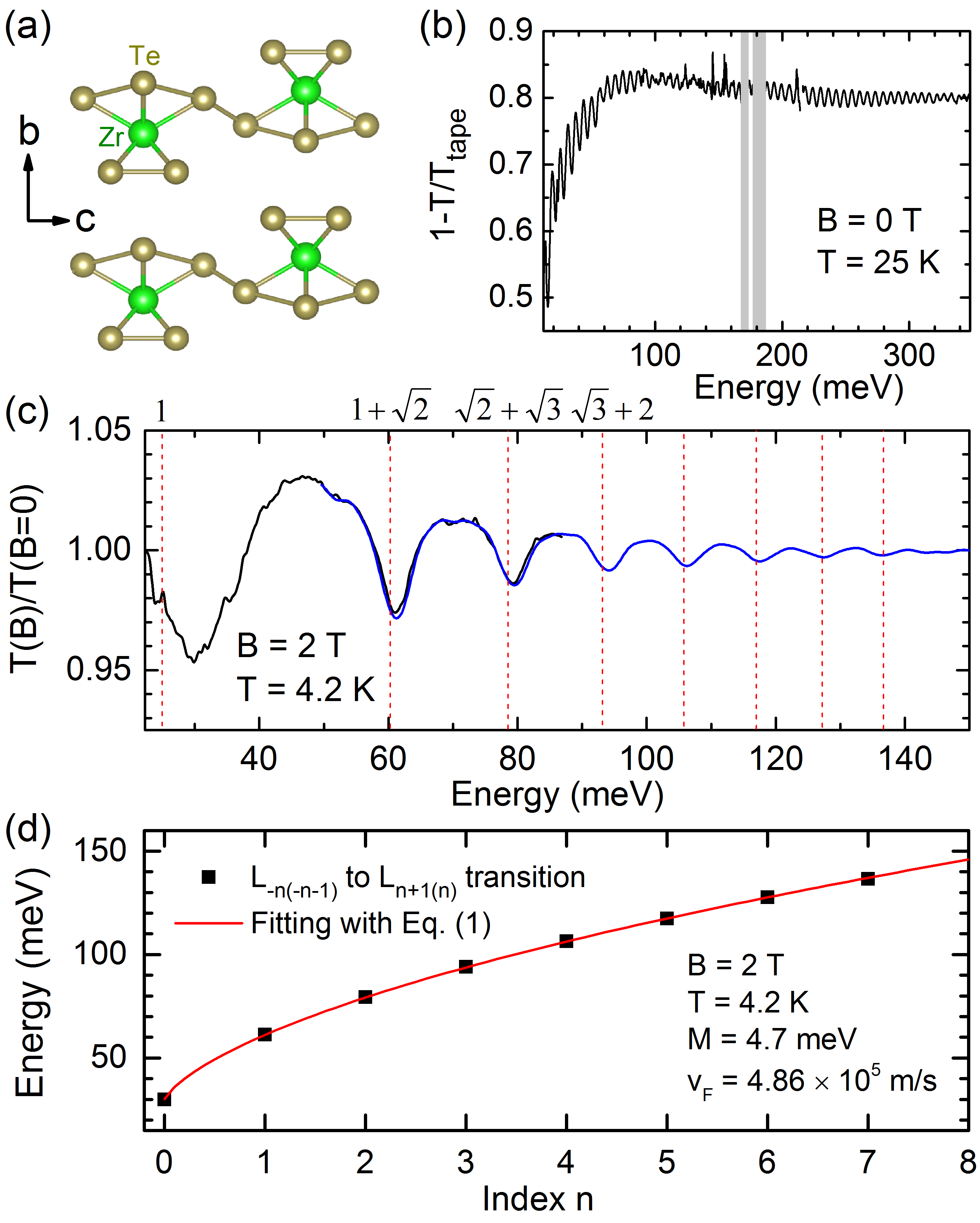}
\caption{(color online) (a) Schematic view of ZrTe$_5$ unit cell along the $a$-axis. The layer stacking direction is along the $b$-axis. (b) Extinction spectrum, $1-T/T_{\text{tape}}$, of ZrTe$_5$/tape composite measured at $B=0$ T and $T=25$ K. The fast oscillations originate from Fabry-P\'{e}rot interference. The gray stripes cover opaque regions due to tape absorption. (c) Normalized transmission spectrum, $T(B)/T(B=0)$, measured at $B=2$ T and $T=4.2$ K. The black and blue curves correspond to the far-IR and the mid-IR spectrum, respectively. The red dash lines mark the expected energies of $L_{-n(-n-1)}\rightarrow L_{n+1(n)}$ transitions for massless Dirac fermions. (d) Extracted LL transition energy from (c) as a function of LL index $n$. The red line shows the best fit to the data using Eq. (1).}
\end{figure}

ZrTe$_5$ single crystals were prepared by the Te-assisted chemical vapor transport (CVT) method \cite{CVT_growth} or molten Te flux growth \cite{Arpes0_GDG}. The crystal has a layered structure with weakly van der Waals coupled layers along the $b$-axis (Fig. 1(a)). By repeatedly exfoliating the material with an IR-transparent Scotch tape, we prepared thin ZrTe$_5$ flakes with the average thickness of about 1 $\mu$m that enables IR transmission/absorption measurements. This method has been proven successful in the previous studies of graphite \cite{MPotemski,RJNicholas} and TI materials such as Bi$_2$Te$_3$ \cite{TItape}. In the main text below, we present the data taken on CVT-grown samples. Similar results were measured with the flux-grown samples, as reported in the Supplemental Material \cite{SM} together with the detailed description of the crystal growth and experimental setup.

In Fig. 1(b), we plot the zero-field extinction spectrum, $1-T/T_{\text{tape}}$, of ZrTe$_5$/tape composite measured at 25 K. Here, the sample spectrum ($T$) is referenced to the transmission through a bare tape ($T_{\text{tape}}$). At low photon energies, the extinction coefficient, and consequently, the absorption ($A$) first increases with energy ($E$) and then becomes spectrally flat at $E>75$ meV. This behavior clearly deviates from the expected linear dependence, $A\propto E$, for 3D Dirac semimetals \cite{Timusk-2013}, and differs our thin flake samples from the thick, opaque samples studied in Refs. \cite{Refl1_NLW,Refl_NLW}, where a 3D massless Dirac semimetal electronic structure was concluded for ZrTe$_5$. Moreover, our data are similar to that observed in graphene \cite{Mak,ZL2}, the best known material system hosting 2D Dirac fermions, for the entire experimental spectral range. Due to its 2D nature, $A=const.$ in mono-, bi-, and multi-layer graphene at high photon energies \cite{flat_theory1,flat_theory2,flat_theory3,flat_theory4}. This 2D Dirac fermion speculation is supported by recent transport studies on ZrTe$_5$ thin flakes \cite{Tp4_WP,TP_XFX,ReflTP_XFX,TP5_XSW}.

To elucidate the electronic structure of ZrTe$_5$ thin flakes, we carry out systematic low-temperature IR transmission measurements in the Faraday geometry in magnetic fields up to $B=16$ T. Figure 1(c) shows a normalized transmission spectrum taken at $B=2$ T featuring a characteristic, graphene-like series of absorption minima. Indeed, the transition energies, which can be readily and accurately determined from the central energy of the absorption line,  can be  be assigned to a series of interband LL transitions from $L_{-n(-n-1)}$ to $L_{n+1(n)}$ with the integer $n$ (or $-n$) being the LL index. The LL spectrum of 2D Dirac fermions such as that in graphene can be described as
\begin{align}
E_n=\alpha\sqrt{2e\hbar v^2_F nB+M^2},
\end{align}
where $E_n$ is the energy of the $n^{th}$ LL, $e$ is the electron charge, $\hbar$ is the reduced Planck's constant, $v_F$ is the Fermi velocity, $M$ is the Dirac mass, and $\alpha=\pm 1$ stands for the conduction and valence bands, respectively. For massless Dirac fermions, $E_n\propto\sqrt{n}$, leading to the characteristic $E_{-n\rightarrow n+1}\propto\sqrt{n}+\sqrt{n+1}$ dependence of optically allowed interband LL transitions $L_{-n}\rightarrow L_{n+1}$ \cite{WAdH,ZJ1}. For massive Dirac fermions ($M\neq0$), however, $E_n$ deviates from a perfect $\sqrt{n}$ dependence \cite{ZL1}. The deviation becomes more pronounced for low-lying LL transitions when $n$ is small. Such a massive Dirac fermion scenario can precisely describe our data at low magnetic fields. The vertical dash lines in Fig. 1(c) indicate transition energies following a model $\sqrt{n}+\sqrt{n+1}$ dependence, with the parameter $v_F$ determined by the highest energy transition ($n=7$). The measured energies of LL transitions exhibit a clear blueshift, particularly for low-lying transitions, suggesting the massive Dirac fermion interpretation. A more quantitative analysis is shown in Fig. 1(d), where the extracted transition energies are plotted as a function of $n$ and fitted with Eq. (1). The best fit to the data gives $M=4.7$ meV (corresponding to a 9.4 meV energy gap) and $v_F=4.86\times10^5$ m/s. The latter is the average Fermi velocity in the \textit{ac} plane of ZrTe$_5$ and its value is consistent with recent transport \cite{Tp1_MLT}, ARPES \cite{Arpes2_XJZ} and IR \cite{Refl1_NLW} measurements.

The observation of a small Dirac mass of $M=4.7$ meV is not a surprise. In theory, the Dirac point in semimetals such as ZrTe$_5$ is composed of two overlapping Weyl points with opposite chirality \cite{gap_theory2,gap_theory1,gap_theory3}. When it lacks symmetry protection, the annihilation of the Weyl points leads to a gap opening at the Dirac point, equivalent to generating a Dirac mass. In addition, the lowest energy transition observed is $L_{0(-1)}\rightarrow L_{1(0)}$, which implies that our sample is in quantum limit. This transition is visible at the magnetic field as low as 0.5 T, corresponding to a Fermi energy $\leq16$ meV. Therefore, our samples are close to the intrinsic limit, suited for magneto-IR spectroscopy studies. Also, we note that the Lorentzian lineshape of LL transitions (Fig. 1(c)) provides another indication in favor of a 2D Dirac fermion picture, as the $k_z$ dispersion in a 3D system would lead to an asymmetric lineshape with abrupt cutoff on the low-energy side \cite{Orlita2}.

\begin{figure}[b]
\includegraphics[width=8.5cm]{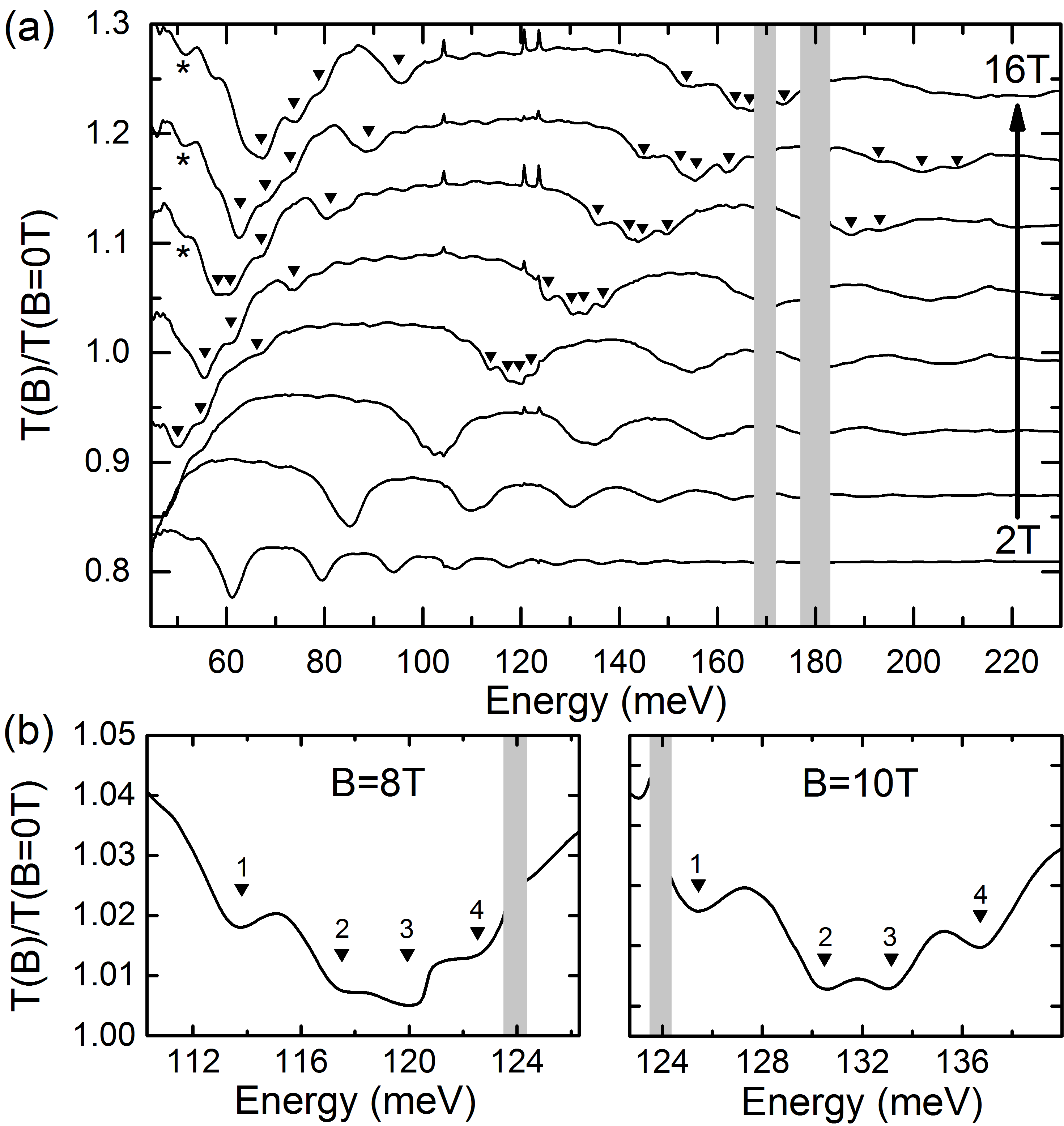}
\caption{(a) Normalized transmission spectra, $T(B)/T(B=0\text{T})$, of ZrTe$_5$/tape composite measured at selected magnetic fields. The down triangles ($\blacktriangledown$) label the splitting of low-lying LL transitions, while the star symbols ($\star$) point to $B$-independent spectral features originating from the normalization process. (b) Zoom-in view of the four-fold splitting of the $L_{-1(-2)}\rightarrow L_{2(1)}$ transition taken at $B=8$ T and 10 T. In all panels, the spectra are offset vertically for clarity and the gray stripes cover opaque regions due to tape absorption.}
\end{figure}

Figure 2(a) illustrates the magnetic field dependence of the LL transitions and their splitting in high magnetic fields, particularly for the three lowest interband transitions: $L_{0(-1)}\rightarrow L_{1(0)}$, $L_{-1(-2)}\rightarrow L_{2(1)}$, and $L_{-2(-3)}\rightarrow L_{3(2)}$. The splitting of the $L_{0(-1)}\rightarrow L_{1(0)}$ transition was previously observed in magneto-IR reflectance measurements \cite{Refl_NLW}, but the proposed interpretation suffers from the requirement of two sets of $g$-factors. In this work, to explore the origins of the splitting, we performed magneto-IR circular polarization resolved measurements using mid-IR  quantum cascade lasers (QCLs) \cite{SM}. Magneto-spectroscopy with circularly polarized light has been successfully employed in the past to reveal details of specific LL transitions in graphite \cite{Toy-Dressel_1977}, and more recently in graphene \cite{Crassee-2011} and in a typical 3D TI Bi$_2$Se$_3$ \cite{Orlita-Bi2Se3-2015}. Here, we focus on the $L_{-1(-2)}\rightarrow L_{2(1)}$ transition, which overlaps well the spectral range of our QCLs. The circular polarization resolved spectra are taken by fixing the light polarization and sweeping the magnetic field in  positive or negative directions, which is equivalent to the use of $\sigma^+$ and $\sigma^-$ polarized light.

\begin{figure}[b]
\includegraphics[width=8.5cm] {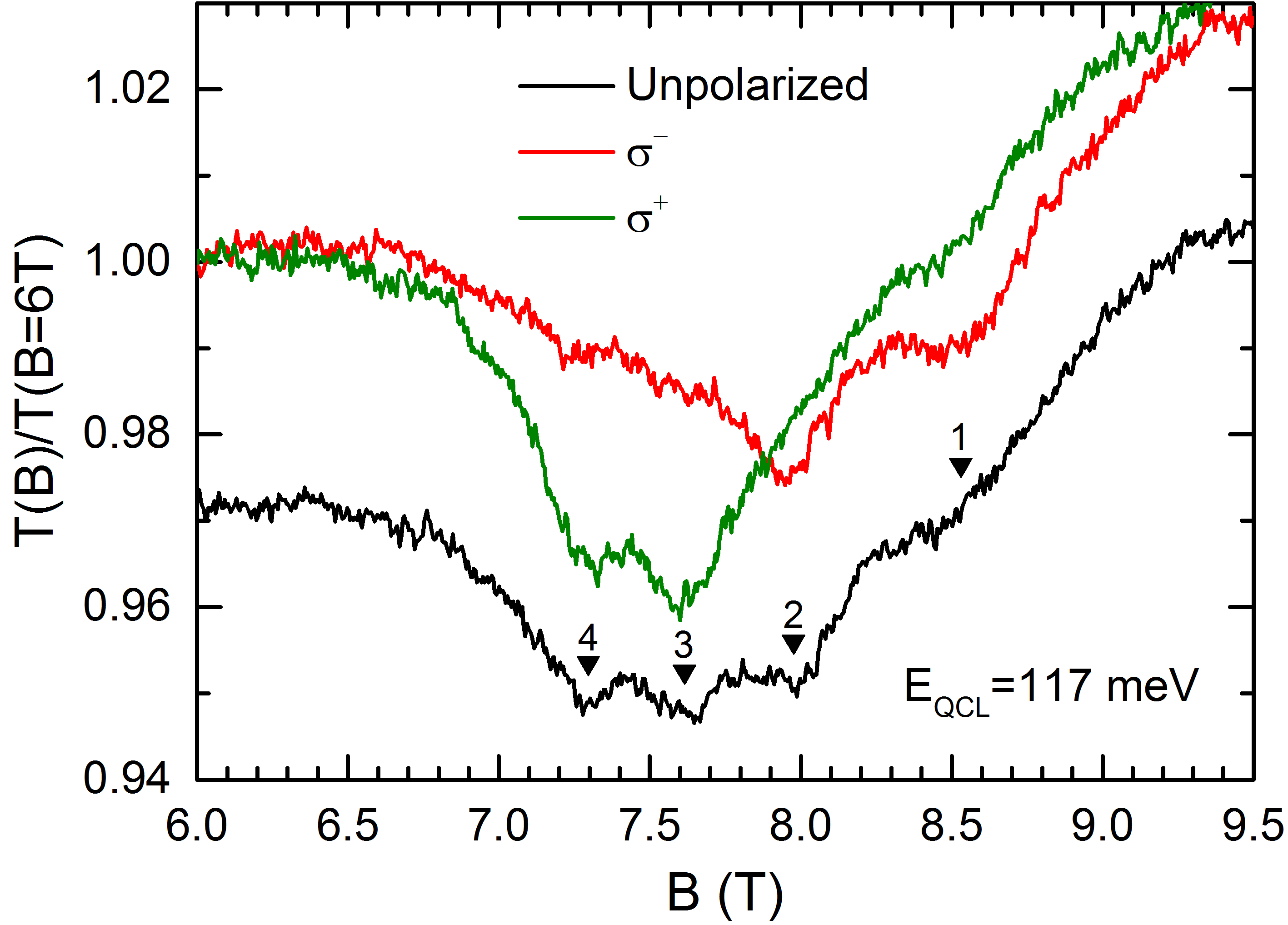}
\caption{(color online) Normalized transmission, $T(B)/T(B=6\text{T})$, through ZrTe$_5$/tape composite as a function of magnetic field with unpolarized (black) and circularly polarized (red and green) incident IR light of 117 meV. The negative magnetic field sweep (green) is flipped to the positive side for the ease of comparison. The four-fold splitting of the $L_{-1(-2)}\rightarrow L_{2(1)}$ transition is labeled by down triangles ($\blacktriangledown$) on the unpolarized data, which is offset vertically for clarity.}
\end{figure}

Figure 3 shows the normalized transmission through ZrTe$_5$/tape composite as a function of magnetic field with the QCL energy fixed at $E_{QCL}=117$ meV. With unpolarized IR light, a four-fold splitting of the $L_{-1(-2)}\rightarrow L_{2(1)}$ transition clearly reproduces that measured at $B=const.$ (Fig. 2(b)). In a circularly polarized configuration, only two of the four split transitions are active in $\sigma^+$ or $\sigma^-$ polarized light. This observation indicates the lifting of the degeneracy between the $L_{-1}\rightarrow L_{2}$ ($\Delta n=1$, $\sigma^+$ active) and $L_{-2}\rightarrow L_{1}$ ($\Delta n=-1$, $\sigma^-$ active) transitions, which can be attributed to an asymmetry between the electron and hole bands.

Next, we show that the remaining two-fold splitting of the $L_{-1}\rightarrow L_{2}$ (or $L_{-2}\rightarrow L_{1}$) transition reflects  the lifting of the spin degeneracy, due to a combined effect of large $g$-factor (Zeeman effect) and finite mass. We begin with an effective Hamiltonian postulated by Bernevig, Hughes, and Zhang \cite{BHZ}
\begin{align*}
H(\bold{k})=\epsilon_0(\bold{k})+
\begin{pmatrix}
    &L(\bold{k}) &Ak_+ &0 &0 \\
    &Ak_- &-L(\bold{k}) &0 &0 \\
    &0 &0 &L(\bold{k}) &-Ak_- \\
    &0 &0 &-Ak_+ &-L(\bold{k})
\end{pmatrix},
\end{align*}
where $\epsilon_0(\bold{k})=C-D(k_x^2+k_y^2)$, $L(\bold{k})=M-\mathscr{B}(k^2_x+k^2_y)$, $k_\pm=k_x \pm i k_y$, and $k^4$ terms neglected \cite{Model1_FZ}. The actual electronic structure is then determined by a set of material parameters: (1) $A=\hbar v_F$, (2) band inversion parameter $\mathscr{B}$, (3) energy offset $C$ (which is set to zero), (4) electron-hole asymmetry parameter $D$, and (5) Dirac mass $M$. In the presence of a magnetic field, one can add the Zeeman term \cite{Molenkamp1,Model2_DS}, $\frac{\mu_B B}{2} 
\begin{pmatrix}
    &g &0 \\
    &0 &-g 
\end{pmatrix}$, where $\mu_B$ is the Bohr magneton, $g =
\begin{pmatrix}
    &g_e &0 \\
    &0 &g_h 
\end{pmatrix}$, and $g_e$($g_h$) are the effective $g$-factors for conduction(valence) bands, and solve the eigenvalue problem for the LL spectrum of massive Dirac fermions in ZrTe$_5$ thin flakes
\begin{widetext}
\begin{align}
E^{\uparrow}_0=M-(D+\mathscr{B})\frac{eB}{\hbar}+\frac{\mu_B g_e}{2}{B} ,   \qquad E^{\downarrow}_0=-M-(D-\mathscr{B})\frac{eB}{\hbar}-\frac{\mu_B g_h}{2}{B}, \qquad &n=0\\
 E^s_{n,\alpha}=-(2Dn+s\mathscr{B})\frac{eB}{\hbar}+s\frac{\mu_B \bar{Z}}{2} B+\alpha  \sqrt{2 A^2 n\frac{e B}{\hbar}+\left[M-(2\mathscr{B}n+sD)\frac{eB}{\hbar}+s\frac{\mu_B \delta Z}{2}{B}\right]^2}.\qquad  &n\neq 0
\end{align}
\end{widetext}
Here, $s=\uparrow\downarrow=\pm 1$ stands for the spin-up and spin-down LLs, $\bar{Z}=\frac{g_e + g_h}{2}$ is the average $g$-factor, and $\delta Z=\frac{g_e - g_h}{2}$. In the low-field limit, Eqs. (2) and (3) reduce to Eq. (1). It should be emphasized that the Zeeman effect alone cannot lift the spin degeneracy of LL transitions even when considering electron-hole asymmetry, $D\neq0$ and $\delta Z\neq0$. This can be seen in Eq. (3), where a finite mass, $M\neq0$ and/or $\mathscr{B}\neq0$ \cite{note2}, is required to distinguish the $\left[ \cdots \right]^2$ term for $s=\pm1$. Therefore, the observed four-fold splitting of low-lying LL transitions provides another evidence of finite mass for the Dirac fermions in ZrTe$_5$ thin flakes. 

\begin{figure}[b]
\includegraphics[width=8.5cm]{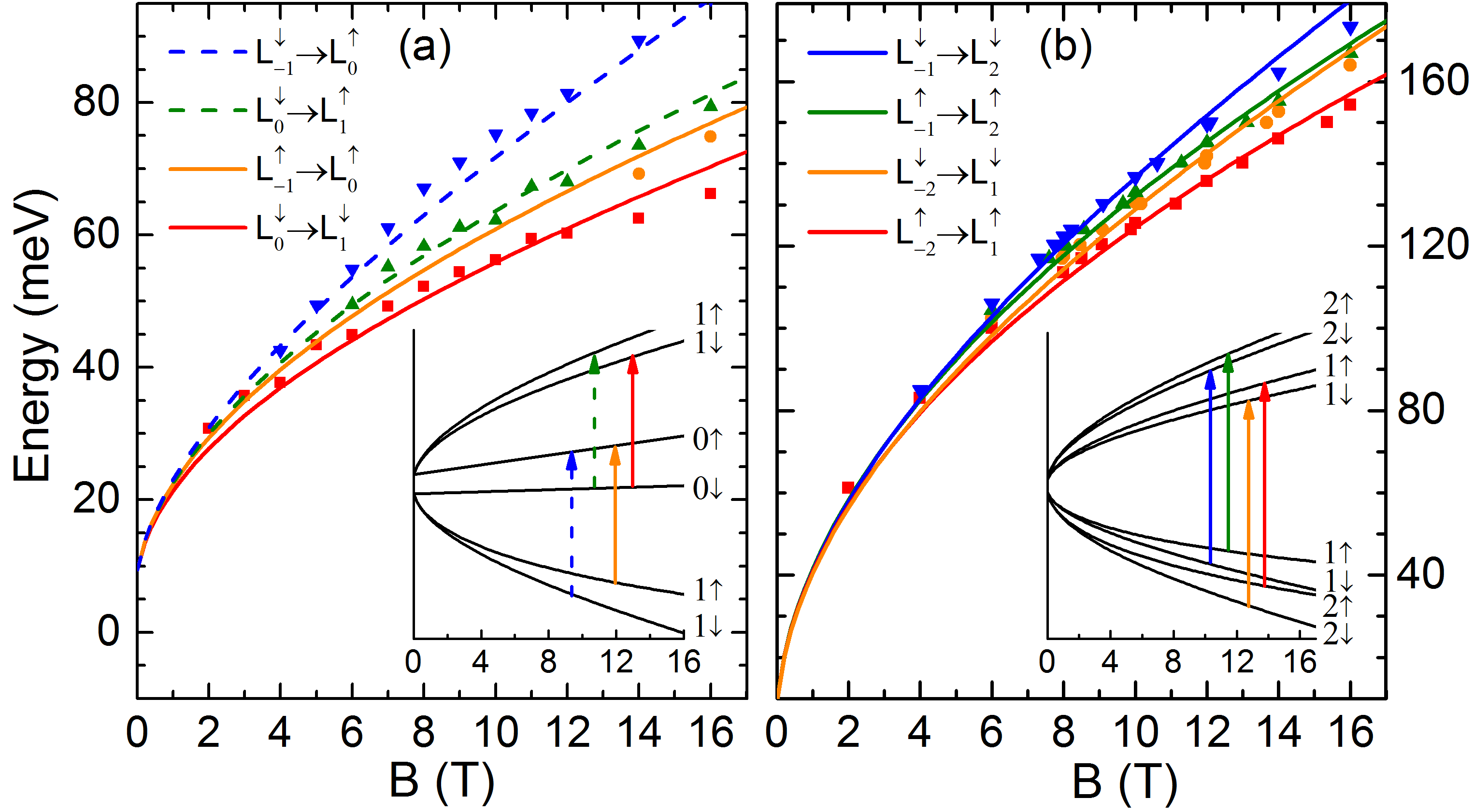}
\caption{(color online) Four-fold splitting of the (a) $L_{0(-1)}\rightarrow L_{1(0)}$ and (b) $L_{-1(-2)}\rightarrow L_{2(1)}$ transitions as a function of magnetic field. Symbols represent the extracted transition energies from both the broad-band measurements at constant magnetic field and the QCL-based measurements at constant photon energy \cite{SM}. The color-coded lines show best fits to the data using Eqs. (2) and (3). The corresponding transitions are illustrated in the insets with the same color code. The solid and dash lines denote the spin-conserved strong transitions and the spin-flipped weak transitions, respectively.}
\end{figure}

Figure 4(b) shows the four-fold splitting of the $L_{-1(-2)}\rightarrow L_{2(1)}$ transition as a function of magnetic field and the color-coded lines are best fits to the data using Eq. (3). The associated fitting parameters are $v_F=4.65 \times 10^5$ m/s, $\mathscr{B}=341$ meV nm$^2$, $D=-126$ meV nm$^2$, $M=4.71$ meV, $g_e=24.3$, and $g_h=7.5$. Here, $M$ is consistent with that obtained from Fig. 1(d) while $v_F$ is $\sim$4\% smaller, and $D<0$ implies a steeper conduction band than the valence band. Interestingly, we notice that a smaller $v_F$ is also needed to better describe the high-field data in Ref. \cite{Molenkamp1}.

To further validate our model, we checked if the above parameters allow to describe other split transitions, and found a very good agreement for the $L_{0(-1)}\rightarrow L_{1(0)}$ transition (Fig. 4(a)). Due to the strong spin-orbit coupling in ZrTe$_5$, spin-flipped LL transitions ($L^{\downarrow}_{0}\rightarrow L^{\uparrow}_{1}$ and $L^{\downarrow}_{-1}\rightarrow L^{\uparrow}_{0}$) are allowed and assigned to the two relatively weak high-energy modes of the $L_{0(-1)}\rightarrow L_{1(0)}$ splitting. The four-fold splitting of the $L_{-1(-2)}\rightarrow L_{2(1)}$ transition, on the other hand, is related to the four strong spin-conserved modes, as the associated spin-flipped modes are expected to be very weak.

Lastly, our model predicts that additional splitting of the $L_{0(-1)}\rightarrow L_{1(0)}$ transition into $L^{\uparrow}_{0}\rightarrow L^{\uparrow}_{1}$ ($L^{\downarrow}_{-1}\rightarrow L^{\downarrow}_{0}$) may occur at lower energies due to the presence of a small amount of electron (hole) doping. Quantitative study of this mode, however, is hindered by a field-independent spectral feature at $\sim$52 meV (labeled by star symbol in Fig. 2(a)) and thus not pursued in this work.

In conclusion, we have performed IR transmission measurements on exfoliated ZrTe$_5$ near the intrinsic limit. The electronic structure of ZrTe$_5$ thin crystals is found to be 2D-like and support a Dirac semimetal interpretation but with a small relativistic mass (or gap). High-field magneto-spectroscopy measurements reveal a four-fold splitting of low-lying LL transitions and circular polarization resolved measurements show that two-fold comes from breaking the electron-hole symmetry while the other two-fold is caused by lifting the spin degeneracy. The magnetic field dependence of the splitting can be fully described using the Bernevig-Hughes-Zhang effective Hamiltonian model.  

Note: During the preparation of this manuscript, we became aware of another IR transmission study of ZrTe$_5$ thin flake \cite{Chen_PNAS}.

We thank Kun Yang and Markus Kindermann for helpful discussions. This work was primarily supported by the DOE (Grant No. DE-FG02-07ER46451). The CVT crystal growth at UT was supported by the NSF (Grant No. DMR-1350002). The flux crystal growth and IR measurements were performed at the National High Magnetic Field Laboratory (NHMFL), which is supported by the NSF Cooperative Agreement No. DMR-1157490 and the State of Florida. Z.J. acknowledges support from the NHMFL Visiting Scientist Program.


\begin{thebibliography}{99}
\bibitem{note2}The parameter $\mathscr{B}$ can also be viewed as a field-induced mass.
\end{thebibliography}
\end{document}


\title{Supplementary Material: Landau level spectroscopy of massive Dirac fermions in single-crystalline ZrTe$_5$ thin flakes}

\author{Y. Jiang}
\affiliation{School of Physics, Georgia Institute of Technology, Atlanta, Georgia 30332}
\author{Z. L. Dun}
\affiliation{Department of Physics and Astronomy, University of Tennessee, Knoxville, Tennessee 37996}
\author{H. D. Zhou}
\affiliation{Department of Physics and Astronomy, University of Tennessee, Knoxville, Tennessee 37996}
\author{Z. Lu}
\affiliation{National High Magnetic Field Laboratory, Tallahassee, Florida 32310}
\affiliation{Department of Physics, Florida State University, Tallahassee, Florida 32306}
\author{K.-W. Chen}
\affiliation{National High Magnetic Field Laboratory, Tallahassee, Florida 32310}
\affiliation{Department of Physics, Florida State University, Tallahassee, Florida 32306}
\author{S. Moon}
\affiliation{National High Magnetic Field Laboratory, Tallahassee, Florida 32310}
\affiliation{Department of Physics, Florida State University, Tallahassee, Florida 32306}
\author{T. Besara}
\affiliation{National High Magnetic Field Laboratory, Tallahassee, Florida 32310}
\author{T. M. Siegrist}
\affiliation{National High Magnetic Field Laboratory, Tallahassee, Florida 32310}
\affiliation{Department of Chemical and Biomedical Engineering, Florida A\&M University-Florida State University (FAMU-FSU)\\ College of Engineering, Florida State University, Tallahassee, Florida 32310}
\author{R. E. Baumbach}
\affiliation{National High Magnetic Field Laboratory, Tallahassee, Florida 32310}
\author{D. Smirnov}
\affiliation{National High Magnetic Field Laboratory, Tallahassee, Florida 32310}
\author{Z. Jiang}
\email{zhigang.jiang@physics.gatech.edu}
\affiliation{School of Physics, Georgia Institute of Technology, Atlanta, Georgia 30332}
\date{\today}

\maketitle
\section{Single Crystal Growth}
Single-crystal ZrTe$_5$ samples were synthesized by both the Te-assisted chemical vapor transport (CVT) method \cite{CVT_growth} and the molten Te flux growth \cite{Arpes0_GDG}. The CVT growth started with polycrystalline ZrTe$_5$, which was prepared by reacting appropriate ratio of Zr and Te in an evacuated quartz tube at 450 $^{\circ}$C for one week. Next, 2 g of polycrystalline ZrTe$_5$ along with transport agent (100 mg Te) were sealed in a quartz tube and placed horizontally in a tube furnace. The sample (source) was placed at the center of the furnace and heated up to 530 $^{\circ}$C at a rate of 60 $^{\circ}$C/hour. The growth zone (sink), which is 12 cm away from the center, was measured to be at a temperature of 450 $^{\circ}$C.

In the flux growth, elemental Zr and Te were mixed in the molar ratio 1:400 and sealed under vacuum in a quartz tube. The mixture was heated at a rate of 50 $^{\circ}$C/hour to 900 $^{\circ}$C, held at this temperature for 72 hours, and then slowly cooled at a rate of 3 $^{\circ}$C/hour to 445 $^{\circ}$C, followed by re-melting the small crystals between 445 and 505 $^{\circ}$C (by rapid heating at 60 $^{\circ}$C/hour and slow cooling at 2 $^{\circ}$C/hour for four times). Finally, the crystals were separated from the flux by centrifuging at 445 $^{\circ}$C.

To remove excess Te on the sample surface, we first seal the ZrTe$_5$ crystals back in an evacuated quartz tube, and then place it in a tube furnace that can create a temperature gradient. The hot and cold zone temperature of the furnace is 400 $^{\circ}$C and 350 $^{\circ}$C, respectively. The crystals are placed in the hot zone for 24 hours, while the excess Te is transported to the cold zone.

Room-temperature X-ray diffraction measurements were performed on both types of samples. Similar lattice constants were obtained, with $a=3.985\ \text{\AA}$, $b=14.526\ \text{\AA}$, and $c=13.716\ \text{\AA}$ for CVT-grown samples and $a=3.988\ \text{\AA}$, $b=14.505\ \text{\AA}$, and $c=13.707\ \text{\AA}$ for flux-grown samples. These values are consistent with that reported in the literature \cite{Arpes0_GDG,ARST_SHP}.

\section{Magneto-infrared Spectroscopy Setups}
Broad-band magneto-infrared spectroscopy measurements were performed using a Bruker 80v Fourier-transform infrared (IR) spectrometer. The (unpolarized) IR light from a mercury lamp (far-IR) or a globar source (mid-IR) was delivered to the sample through evacuated light pipes. The sample was placed at the center of a 17.5 T superconducting magnet. During the measurement, IR transmission spectra were taken at selected magnetic fields and normalized to the zero-field spectrum.

Circular polarization resolved magneto-IR measurements are based on wavelength tunable quantum cascade lasers (QCLs) covering 807-1613 cm$^{-1}$ spectral range. Schematics of the experimental setup are shown in Fig. S1. The circular polarization was generated through a set of IR polarizing components and the IR beam was focused on the sample using a long parabolic cone. During the measurement, the QCL was fixed at a constant photon energy, while the magnet field was swept in both positive and negative directions.

For both setups, the transmitted light was collected by a composite Si bolometer mounted behind the sample at liquid helium temperature. For the latter, to reduce excess noise the light was modulated with a chopper and detected by a lock-in. All measurements were performed in the Faraday configuration with the $b$-axis of the sample parallel to the magnetic field.

\begin{figure}[h]
\includegraphics[width=14cm]{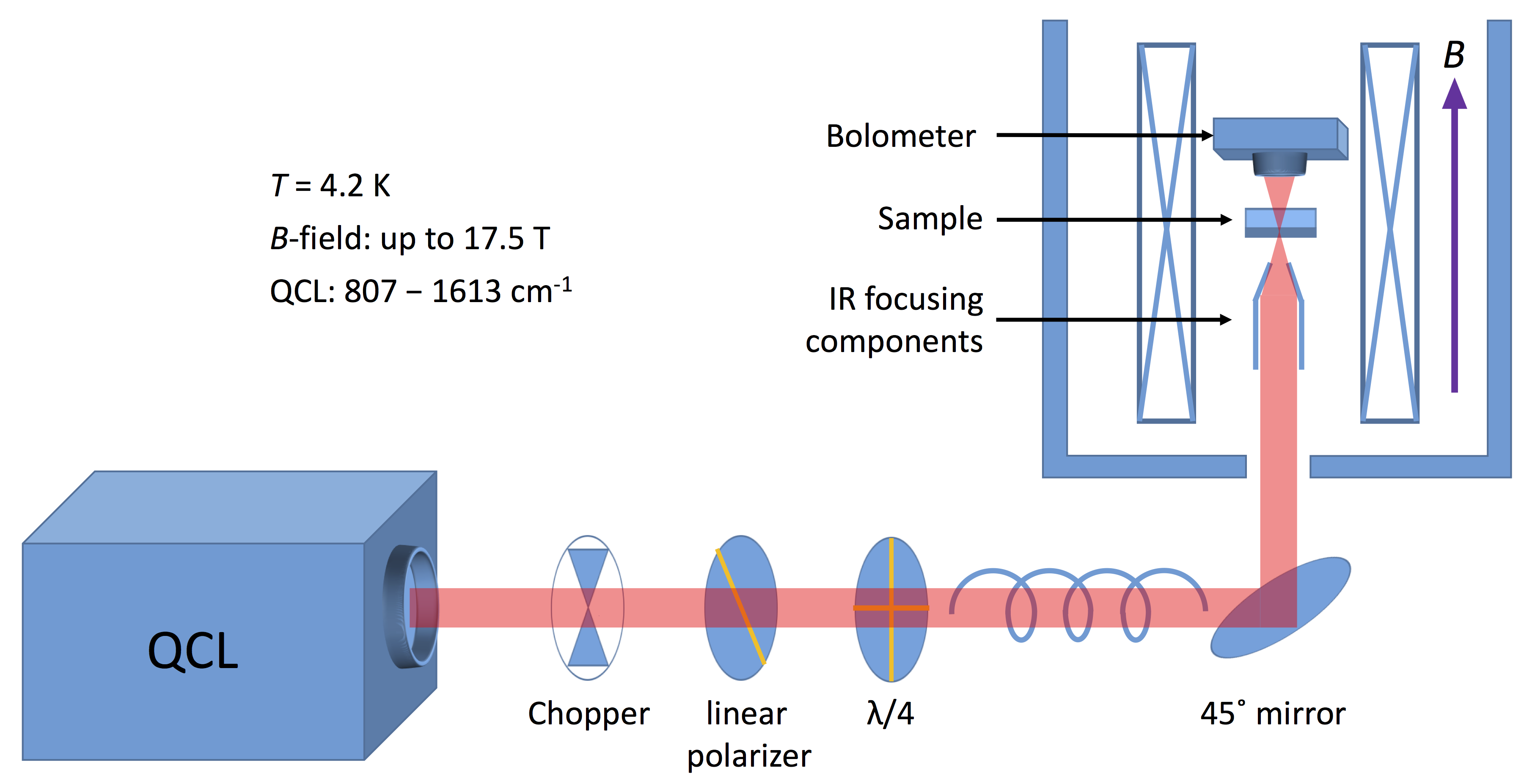}
\caption{(color online) Schematics of QCL-based setup for circular polarization resolved mid-IR magneto-spectroscopy.}
\end{figure}

\section{Additional Experimental Data on a Second CVT-grown Sample}
Figure S2 shows the normalized transmission spectra of a second CVT-grown ZrTe$_5$ sample at selected magnetic fields. It is consistent with Fig. 2 of the main text.

\begin{figure}[t]
\includegraphics[width=14cm]{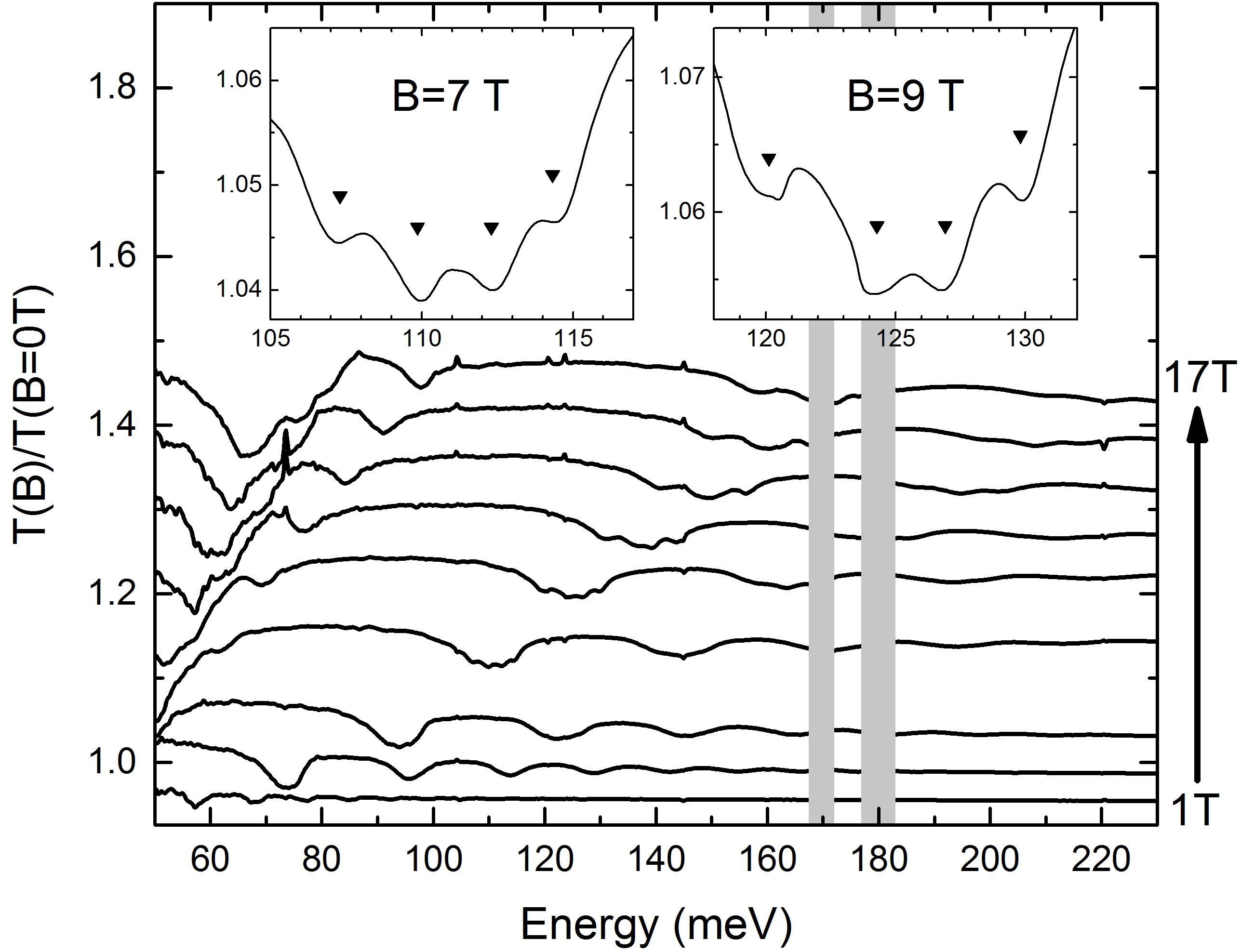}
\caption{Normalized transmission spectra, $T(B)/T(B=0\text{T})$, of a second CVT-grown ZrTe$_5$ sample at selected magnetic fields. Inset: Zoom-in view of the four-fold splitting of the $L_{-1(-2)}\rightarrow L_{2(1)}$ transition, labeled by down triangles ($\blacktriangledown$), taken at $B=7$ T and 9 T. The spectra are offset vertically for clarity and the gray stripes cover opaque regions due to tape absorption.}
\end{figure}

\section{Complementary Experimental Data on Flux-grown Samples}
Figure S3 shows the zero-field and low-field data taken on a flux-grown ZrTe$_5$ sample, and Fig. S4 shows the high-field data and field dependence. These results are consistent with that in Figs. 1 and 2 of the main text on a CVT-grown sample, although the four-fold splitting of low-lying Landau level (LL) transitions is not fully developed.

\begin{figure}[t]
\includegraphics[width=14cm]{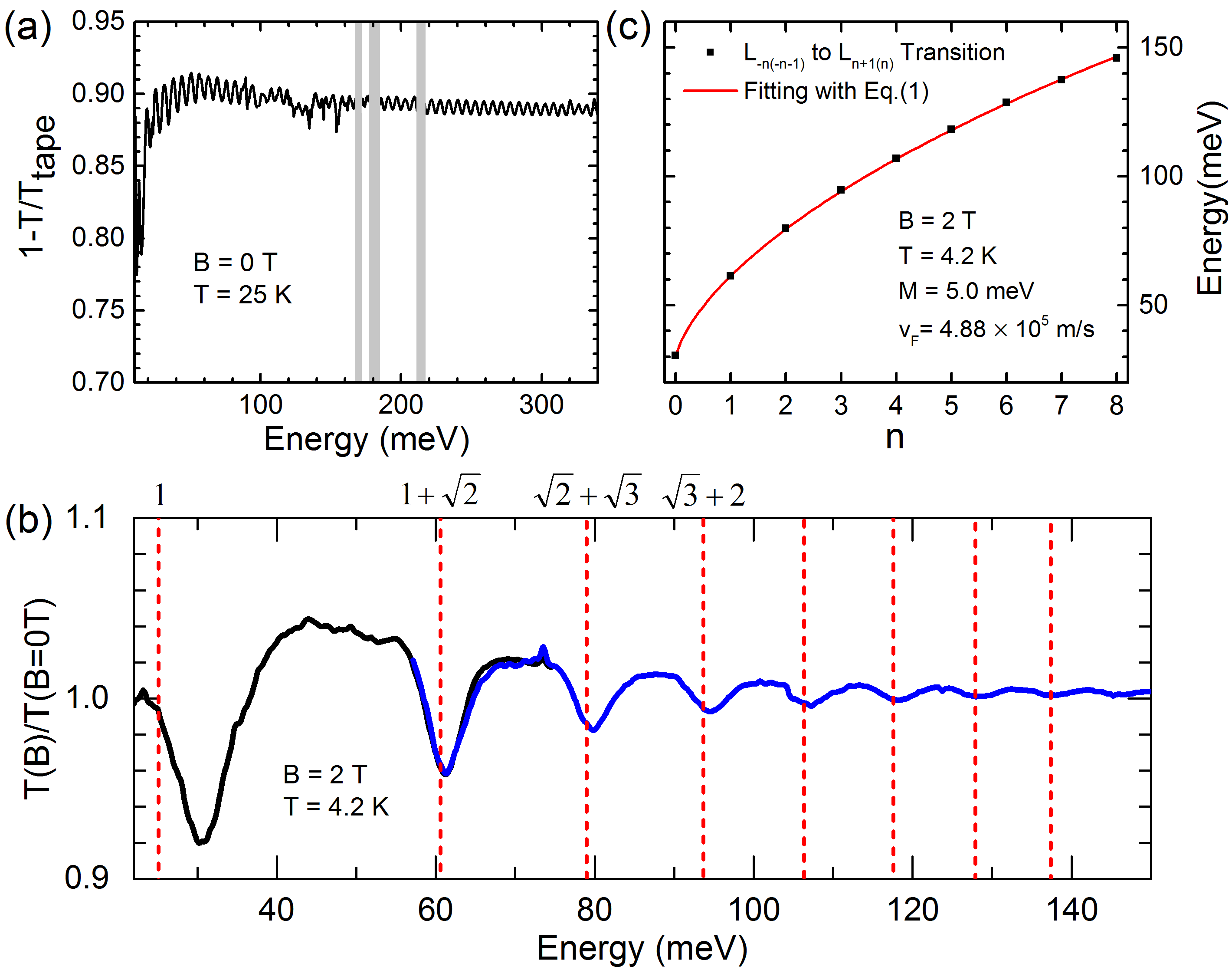}
\caption{(color online) (a) Extinction spectrum, $1-T/T_{\text{tape}}$, of flux-grown ZrTe$_5$/tape composite measured at $B=0$ T and $T=25$ K. The fast oscillations originate from Fabry-P\'{e}rot interference. The gray stripes cover opaque regions due to tape absorption. (b) Normalized transmission spectrum, $T(B)/T(B=0)$, measured at $B=2$ T and $T=4.2$ K. The black and blue curves correspond to the far-IR and the mid-IR spectrum, respectively. The red dash lines mark the expected energies of $L_{-n(-n-1)}\rightarrow L_{n+1(n)}$ transitions for massless Dirac fermions (with the integer $n$ being the LL index). (c) Extracted LL transition energy from (b) as a function of $n$. The red line shows the best fit to the data using Eq. (1) of the main text.}
\end{figure}

\begin{figure}[t]
\includegraphics[width=14cm]{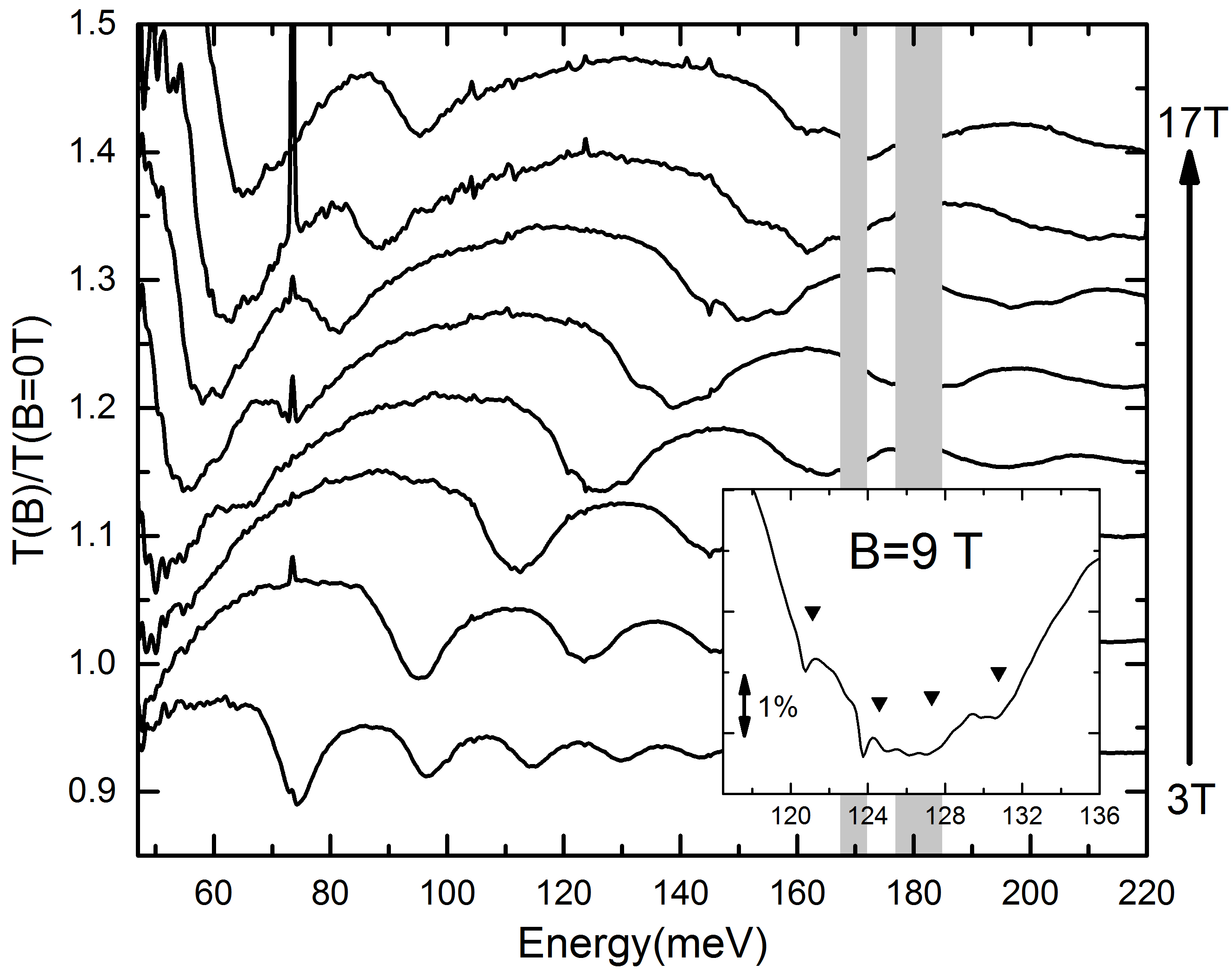}
\caption{Normalized transmission spectra, $T(B)/T(B=0\text{T})$, of flux-grown ZrTe$_5$/tape composite at selected magnetic fields. Inset: Zoom-in view of the four-fold splitting of the $L_{-1(-2)}\rightarrow L_{2(1)}$ transition, labeled by down triangles ($\blacktriangledown$), taken at $B=9$ T. The spectra are offset vertically for clarity and the gray stripes cover opaque regions due to tape absorption.}
\end{figure}

\section{Additional Circular Polarization Resolved Data}
Figure S5 shows additional circular polarization resolved data at different laser energies on CVT-grown ZrTe$_5$ sample. It supports Fig. 3 of the main text. 

\begin{figure}[h]
\includegraphics[width=14cm]{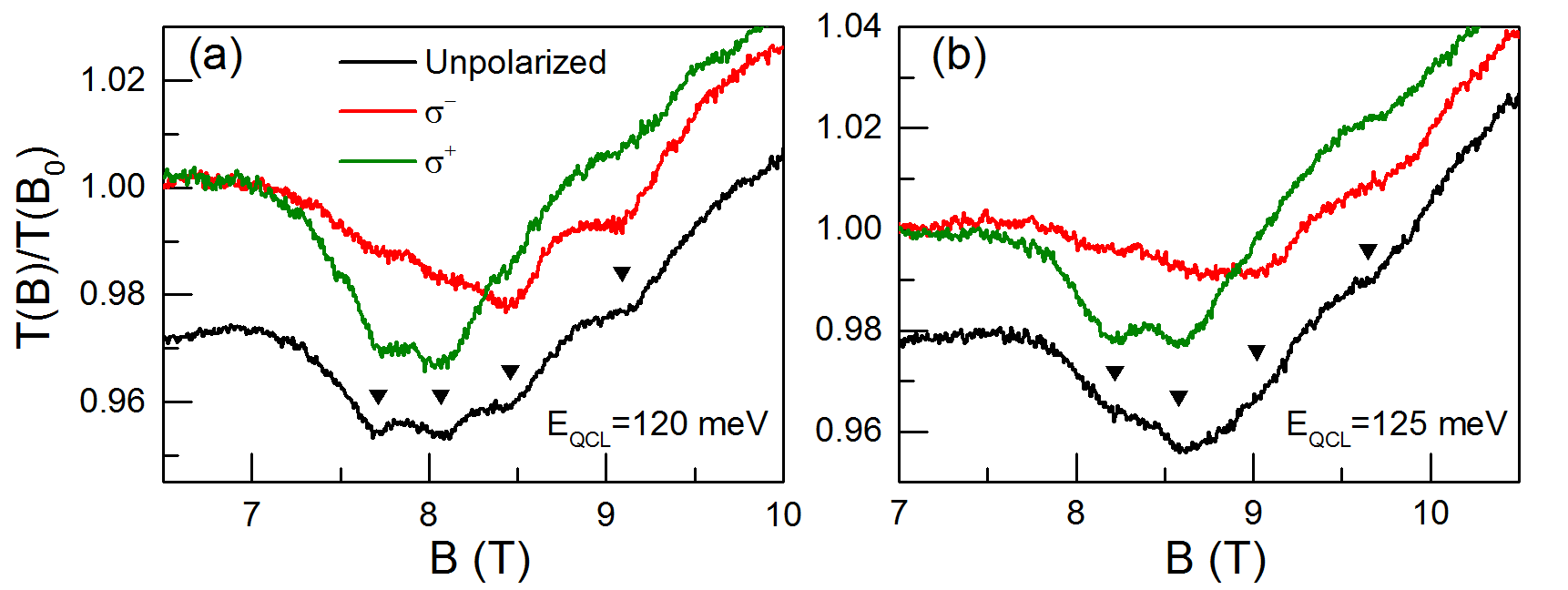}
\caption{(color online) Normalized transmission, $T(B)/T(B_0)$, through ZrTe$_5$/tape composite as a function of magnetic field with unpolarized (black) and circularly polarized (red and green) incident IR light. (a) $E_{QCL}=120$ meV and normalized to $B_0=6$ T and (b) $E_{QCL}=125$ meV and normalized to $B_0=7$ T. The negative magnetic field sweep (green) is flipped to the positive side for the ease of comparison. The four-fold splitting of the $L_{-1(-2)}\rightarrow L_{2(1)}$ transition is labeled by down triangles ($\blacktriangledown$) on the unpolarized data, which is offset vertically for clarity.}
\end{figure}

\section{Extended Effective Hamiltonian Model}
Lastly, we show that Eqs. (2) and (3) in the main text also work for three-dimensional (3D) Dirac semimetals when taking the $k_z=0$ limit. We start with the effective Hamiltonian of Ref. \cite{Refl_NLW} constructed using the following bases: $\ket{+,\uparrow}, \ket{+,\downarrow},\ket{-,\uparrow}, \ket{-,\downarrow}$, where $\pm$ labels the orbitals (or Kramer pairs) and $\uparrow,\downarrow$ denotes the spin. Then, we add the symmetry-allowed $k^2$ terms and obtain
\begin{align*}
H(\bold{k})=(M-\mathscr{B'} v_F^2k^2) \tau^z I_2+\hbar(v_{F,x}k_x\tau^x\sigma^z-v_{F,y}k_y\tau^y I_2)-D' v_F^2k^2 I_4+(\bar{Z}' I_2+\delta Z' \tau^z) \sigma^z,
\end{align*}
where $M$ is the Dirac mass, $\mathscr{B'}$ is the band inversion parameter, $D'$ is the electron-hole asymmetry parameter, $\hbar$ is the reduced Planck's constant, $v_F$ is the Fermi velocity, $v_F^2k^2=v_{F,x}^2k_x^2+v_{F,y}^2k_y^2$, $k_z=0$,  $I_2$ is a 2 by 2 unit matrix,  $I_4$ is a 4 by 4 unit matrix, $\tau^z I_2=
\begin{pmatrix}
    &I_2 &0 \\
    &0 &-I_2 
\end{pmatrix}$, 
$\tau^x\sigma^z=
\begin{pmatrix}
    &0 &\sigma^z \\
    &\sigma^z &0 
\end{pmatrix}$, 
$\tau^y I_2=
\begin{pmatrix}
    &0 &-i I_2 \\
    &i I_2 &0 
\end{pmatrix}$, 
$I_2\sigma^z=
\begin{pmatrix}
    &\sigma^z &0 \\
    &0 &\sigma^z
\end{pmatrix}$, 
$\tau^z \sigma^z=
\begin{pmatrix}
    &\sigma^z &0 \\
    &0 &-\sigma^z 
\end{pmatrix}$, and $\sigma^z$ is a Pauli matrix. In addition, the Zeeman effect is considered with $\bar{Z}'=\mu_B B \bar{g}/2$ and $\delta Z'=\mu_B B \delta g/2$, where $\bar{g}=\frac{g_+ + g_-}{2}$, $\delta{g}=\frac{g_+ - g_-}{2}$, $\mu_B$ is the Bohr magneton, $g_+$ and $g_-$ denote the $g$-factors for the $+$ and $-$ orbitals, respectively. Now the Hamiltonian reads
\[
H= 
\begin{pmatrix}
&H_0 &H_{01}\\
&H_{10} &H_1\\
\end{pmatrix},\]
where
\[
H_0=
\begin{pmatrix}
    &(M-\mathscr{B'}v_F^2k^2)-D' v_F^2k^2+\bar{Z}'+\delta Z'   & 0 \\
    &0                                                       &(M-\mathscr{B'}v_F^2k^2)-D' v_F^2k^2-\bar{Z}'-\delta Z'
\end{pmatrix},
\]
\[
 H_1=
\begin{pmatrix}
&-(M-\mathscr{B'}v_F^2k^2)-D' v_F^2k^2+\bar{Z}'-\delta Z'        &0\\
 &0                                                   &-(M-\mathscr{B'}v_F^2k^2)-D' v_F^2k^2-\bar{Z}'+\delta Z'
\end{pmatrix},
\]
and
\[
H_{01}=H_{10}^\dagger=
\begin{pmatrix}
&\hbar v_{F,x} k_x+i\hbar v_{F,y}k_y &0\\
&0                                     &-\hbar v_{F,x}k_x+i\hbar v_{F,y}k_y\\
\end{pmatrix}.
\]

To obtain the LL energy, we employ the Landau-gauge vector potential $\bold{A}=(-By,0,0)$ and define 
$\Delta=\sqrt{|2\hbar v_{F,x} v_{F,y} e B|}$ and $\hat{b}=\frac{(v_{F,x}\hat{P_x}-iv_{F,y}\hat{P_y})}{\Delta}$ (where $e$ is the electron charge and $\hat{P_x}$ and $\hat{P_y}$ are momentum operators). Then, $[\hat{b},\hat{b}^{\dagger}]=1$ and $v_F^2k^2=(\hat{b}^{\dagger}\hat{b}+\frac{1}{2})\Delta^2=(n+\frac{1}{2})\Delta^2$. The Hamiltonian in a magnetic field becomes
\[
H_0=
\begin{pmatrix}
    &M-(\mathscr{B'}+D')(n-\frac{1}{2})\Delta^2+\bar{Z}'+\delta Z'   & 0 \\
    &0                                                       &M-(\mathscr{B'}+D')(n+\frac{1}{2})\Delta^2-\bar{Z}'-\delta Z'
\end{pmatrix},
\]
\[
 H_1=
\begin{pmatrix}
 &-M+(\mathscr{B'}-D')(n+\frac{1}{2})\Delta^2+\bar{Z}'-\delta Z'   & 0 \\
    &0                                                       &-M+(\mathscr{B'}-D')(n-\frac{1}{2})\Delta^2-\bar{Z}'+\delta Z'
\end{pmatrix},
\]
and
\[
H_{01}=H_{10}^{\dagger}
=
\begin{pmatrix}
&\Delta \sqrt{n} &0\\
&0                                     &-\Delta \sqrt{n}\\
\end{pmatrix}.
\]
The corresponding LL energies are
\begin{align}
E_0^{\uparrow}&=M-\frac{\mathscr{B'}+D'}{2}\Delta^2+\bar{Z}'+\delta Z',\\
E_0^{\downarrow}&=-M+\frac{\mathscr{B'}-D'}{2}\Delta^2-\bar{Z}'+\delta Z',
\end{align}
and for $n\neq 0$,
\begin{align}
 E^s_{n,\pm}=(-\frac{s\mathscr{B'}}{2}-nD')\Delta^2+s\bar{Z}' &\pm \sqrt{\Delta^2 n+\left[M-(n\mathscr{B'}+\frac{sD'}{2})\Delta^2+s\delta Z'\right]^2},
\end{align}
where $s=\uparrow\downarrow=\pm 1$ stands for the spin-up and spin-down LLs. These solutions are essentially the same as Eqs. (2) and (3) in the main text.